\documentclass{article}

     \usepackage[preprint]{neurips_2020}

\usepackage[utf8]{inputenc} 
\usepackage[T1]{fontenc}    
\usepackage{hyperref}       
\usepackage{url}            
\usepackage{booktabs}       
\usepackage{amsfonts}       
\usepackage{nicefrac}       
\usepackage{microtype}      
\usepackage{enumitem}
\usepackage{verbatim}       

\title{Security and Machine Learning in the Real World}

\author{%
  Ivan Evtimov \\
  University of Washington\thanks{Work done while at Microsoft Research} \\
  \texttt{ie5@cs.washington.edu} \\
  \And
  Weidong Cui \\
  Microsoft Research \\
  \texttt{weidong.cui@microsoft.com} \\
  \And
  Ece Kamar \\
  Microsoft Research \\
  \texttt{eckamar@microsoft.com} \\
  \And
  Emre K\i c\i man \\
  Microsoft Research \\
  \texttt{emrek@microsoft.com} \\
  \And
  Tadayoshi Kohno \\
  University of Washington \\
  \texttt{yoshi@cs.washington.edu} \\
  \And
  Jerry Li \\
  Microsoft Research \\
  \texttt{jerrl@microsoft.com} \\
}

\begin{document}

\maketitle

\begin{abstract}
Machine learning (ML) models deployed in many safety- and business-critical systems are vulnerable to exploitation through adversarial examples. 
A large body of academic research has thoroughly explored the causes of these blind spots, developed sophisticated algorithms for finding them, and proposed a few promising defenses.
A vast majority of these works, however, study standalone neural network models.
In this work, we build on our experience evaluating the security of a machine learning software product deployed on a large scale to broaden the conversation to include a systems security view of these vulnerabilities. 
We describe novel challenges to implementing systems security best practices in software with ML components.
In addition, we propose a list of short-term mitigation suggestions that practitioners deploying machine learning modules can use to secure their systems.
Finally, we outline directions for new research into machine learning attacks and defenses that can serve to advance the state of ML systems security. 

\end{abstract}

\section{Introduction}
\label{submission}
Machine learning (ML) advances in the processing of visual, language, and other digital data signals are being rapidly deployed in real-world systems that have critical societal and business significance.
As with any computer technology deployed at scale or in vital domains, ML systems face motivated adversaries intent on causing undesired behaviors or violating security restrictions.  
Recent research shows that so-called {\em adversarial examples} violate security properties of individual models due to fundamental flaws in deep neural networks.
Such adversarial inputs are concerning because they allow adversaries to fully control the model outputs.
This has spurred much research on this threat, largely focused on standalone computer vision models and adversaries that are restricted to direct, small modifications of the pixel inputs (so-called $\ell_p$-bounded adversarial perturbations).
Multiple cycles of attack-defend iterations on such algorithmic approaches to securing and breaking individual models have led to valuable insights for ML research. 

However, algorithmic defenses have not substantially improved the security of ML systems in practice because the assumptions underlying these defenses are often violated in real-world deployments. 
To achieve ML systems and applications that are secure against adversaries ---critical to implementing trustworthy and reliable AI---
the adversarial ML research community must broaden the security conversation beyond algorithmic defenses per se.

In this work, we present the lessons learned from a full-access exploration of deployed ML systems with millions of users.
This analysis was carried out by a team of researchers with expertise in AI safety and robustness, ML theory, large-scale systems reliability, and systems security. 
Based on this experience, we believe that research on robust and secure AI must expand its focus to address the real-world problems that present themselves in practice.
While we do not believe those recommendations are exhaustive, we believe they would benefit other researchers in pursuing more secure ML systems.
We organize our observations and recommendations along two axes.

\paragraph{Rethinking Adversarial ML Research Through the Lens of Systems Security}
First, we join a growing chorus of academic work (\cite{sharif2016accessorize, athalye2017synthesizing, eykholt2018robust, gilmer2018motivating}) in observing that adversarial examples of the kind most commonly studied in the literature are not a significant concern for real-world deployments. 
Most relevant adversaries are incapable of introducing small-magnitude modifications in model inputs, nor can they be restricted to making only these modifications, as most published work assumes.
Therefore, all known defenses, broken or stable, do not provide guarantees against the attacks most likely to be mounted in deployed systems.

Rather, the design of secure machine learning systems should be directly thought of as a systems security question.  Systems security---the study of how to secure the behavior of complex software systems against malicious attacks---has developed a set of principles that are now commonly used to build secure systems with some insecure components.  
When viewed from the lens of systems security first principles, we see that many of the important technical questions drastically change. 
In Section~\ref{sec:challenges}, we describe how this reframing leads us to consider issues including end-to-end ML security, robust input modalities, and adaptive defenses.

\paragraph{Retooling Systems Security for Adversarial Machine Learning}
Second, while adversarial machine learning can learn much from systems security principles, we find that many current practices are difficult to implement, or simply incompatible with state-of-the-art machine learning.  
For example, even basic countermeasures such as checking for invalid inputs to a system and updating software to patch security problems encounter significant challenges in the context of modern ML methods.

The field of adversarial machine learning must resolve these issues. 
In some cases, this requires retooling security practices to be relevant to AI-driven software; in others, our basic understanding of machine learning must be advanced.
In Section~\ref{sec:retooling}, we describe a number of ways in which current practices for securing AI either fall short of or are at odds with well-established security principles.
We also discuss short-term mitigations to these issues and outline where fundamental advances in research are needed to bridge the gap.

\section{Background}
A wide array of work has explored the security of machine learning algorithms over the past several years. 
We can divide the research most relevant to this paper's focus into two categories: attack and defense research on individual models and high-level research seeking to systematize and critique the trends in the previous category. 

\subsection{Algorithmic Attacks and Defenses}

Machine learning models  have long been studied as subjects of adversarial pressure~\citep{dalvi2004adversarial,lowd2005adversarial,lowd2005good}.
This paper concentrates on deep learning models since the important advances they enable in computer vision, natural language processing, and other fields are being widely deployed in real systems facing adversarial pressure. 
The academic study of ``adversarial examples'' against neural networks traces its origin to \cite{szegedy2013intriguing}, which observed that minor perturbations to the inputs of vision models can shift their outputs in unexpectedly major ways. 
Since then, many researchers have found new ways to generate adversarial inputs and make models resistant to such blind spots. 
Most notably, \cite{carlini2017towards} developed a gradient-based optimization method that defeated state-of-the-art defenses proposed at the time and has served as the basis for developing  many attack algorithms. 
This led to the establishment of a ``standard'' threat model for neural networks that is widely deployed in the ML literature. 
Namely, most research considers a network to be ``robust'' if it can be guaranteed -- empirically or theoretically -- that a certain percentage of adversarially modified test set members are correctly classified.
These adversarial modifications must be within an $\epsilon$ distance (usually an $\ell_p$ norm of some sort) of the original, clean examples. 
This metric is known as \textit{robust accuracy}.
Within the bounds of this model, promising approaches to making neural networks more robust have been proposed, such as \cite{madry2017towards}, \cite{lecuyer2019certified}, and \cite{wong2017provable}.

Nevertheless, it remains unclear how this threat model applies in real deployments of vision models or to ML models used in other domains.
Some research moves closer to answering this question.
\cite{sharif2016accessorize} demonstrate that physical objects (e.g., glasses frames) can be designed adversarially to mislead face recognition models; \cite{eykholt2018robust} show that adversarially crafted stickers can cause errors in road sign and image recognition models; and \cite{cao2019adversarial} demonstrate that depth-based networks operating on LiDAR data are also vulnerable. 
Beyond images, \cite{carlini2018audio} craft adversarial sounds that mislead speech transcription models, and \cite{gleave2019adversarial} show that even the input state space of reinforcement learning algorithms can be adversarially modified to tamper with their operation. 

All this work illuminates what is possible under different assumptions of adversarial power and has produced fundamental advances that take us closer to secure models. We offer a new perspective based on our experience evaluating the security of deployed end-to-end systems incorporating deep learning.
We find that generating adversarial examples is not the only method available to attackers. In fact, much more unsophisticated attacks are currently more probable and just as hard to defend against. This realization poses both new challenges that have not yet been considered in the literature and presents under-researched opportunities to reason about security at the system level. 

\subsection{Security and Machine Learning}

In another category of related work, researchers have stepped back to evaluate the threat models and assumptions of the attacks and defense literature. 
Before the emergence of deep learning models and adversarial examples, \cite{barreno2010security} studied how machine learning and security interact. 
\cite{papernot2018sok} first systemized the literature on security and privacy as it relates to deep learning.
This was later extended by \cite{papernot2018marauder}, who applied the 10 security principles of \cite{saltzer1975protection} to highlight how machine learning violates them. While some of our findings overlap, we observe a new set of challenges that have not yet been discussed.

One work that takes a similar approach to ours but applies principles of software engineering instead of security is \cite{sculley2014machine}. 
The authors observe that ML violates typical notions of software isolation, introduces costly data dependencies, and encourages the creation of ``spaghetti'' code. Some of these observations also apply to the security of machine learning systems, as we highlight below.

Finally, \cite{gilmer2018motivating} consider what the correct threat model for studying adversarial examples should be. The authors critique the ``perturbation literature'' -- the set of works studying attacks and defenses bounded by $L_p$ norms -- and find that most motivating scenarios for this work do not apply directly to the setting that is being studied. 
They find that no realistic scenario requires imperceptible adversarial examples and that less sophisticated attacks exist for many others. 
 Our experience with evaluating the security of a deployed system incorporating deep neural networks confirms this observation. Adversarial inputs are not limited to adversarial examples. The challenges we highlight are not limited to adversarial patches, and the solutions we suggest apply more broadly. 
We concretize solutions and expand on them in greater detail than \cite{gilmer2018motivating}.

\section{Adversarial Machine Learning Through the Lens of Systems Security}
\label{sec:challenges}

We propose focusing the  design of robust real-world ML using systems security questions.
While not a perfect success story, systems security research has managed to decrease the risk of threats to traditional software using a robust toolbox of security best practices.
Based on our direct experiences with our industry's ML systems at scale, two ``first principles'' considerations from systems security can aid us in adding a new focus to adversarial machine learning research:

\paragraph{Threat modeling} The first step to increasing the security of any system is developing a threat model and AI systems are no exception.
While there is no single unified approach to threat modeling and risk management in the computer security field, the process often involves carefully thinking through who might want to abuse the system, what capabilities they have, how they might be able to do it, and what the overall risk of the exploit is. 
This process informs designers and engineers where defense efforts should be focused and what compromises can be anticipated. 
In the discussion that follows, we consider how threat modeling machine learning systems can be better informed.

\paragraph{Work factor analysis}
The goal of systems security is in most cases, not to perfectly secure the system, as this is usually impossible.
Rather, it is to introduce as many barriers to an adversary as possible.
If the resources necessary to attack the system become sufficiently high, it becomes disadvantageous for the adversary to attempt to exploit any weaknesses in the system.
In the seminal paper of~\cite{saltzer1975protection}, this is codified as the principle of \emph{work factor.}
Similarly, when designing defenses for robust machine learning, the goal should be to make it as costly as possible for an attacker to maliciously change the behavior of our system, with any attack.  More specifically, ML systems should be designed so that attackers must commit a disproportionately large effort to overcome a small defensive effort.  
This perspective borrowed from systems security yields a number of concrete suggestions to designers of secure ML systems.

In the remainder of this section, we describe the implications of threat modeling and work factor analysis on robust ML and short-term mitigations.

\subsection{End-to-end ML Security}
A machine learning model's interaction with other components in deployed systems can be extremely complex. 
This increases the potential for subtle security flaws but also presents additional possible barriers to adversaries.
We propose focusing more attention on these two issues.

First, threat models should consider system-level interactions. 
Most existing studies of attacks on machine learning consider a single model in isolation and demonstrate that its inputs can be manipulated to shift its predictions according to an adversary's goal.
But when a model is part of a bigger system, a multitude of additional security issues present themselves. 
In developing attacks with full access to the systems in our organization, we found that vulnerabilities were often caused by how programmed logic combined outputs of multiple ML models.
These errors are not fixed by simply improving the accuracy of each of the component models.
As a concrete example of one issue that we observed, models often act on outputs of other models.
If one of them fails to stop the processing of invalid inputs, downstream models will be acting on data outside of their valid operational region.
If the inputs are adversarial, this leads to a full system compromise.
If inputs are merely out-of-distributions, downstream models and code start behaving in undefined ways.
This exposes information about the system and aids adversaries in their search for adversarial inputs.
Therefore, a threat model should consider how a failure in one component leads to cascading failures of the system overall.

Second, a work factor analysis of system-level defenses can help to ensure secure behavior even when an individual model may be vulnerable to adversarial examples or other adversarial inputs.
Our full-access explorations revealed that adversaries seldom succeed on their first attempt to compromise a system.
This was true even when they possessed thorough knowledge of every software component and the model's architecture, parameters, and configuration.
This suggests that systems should stop feeding inputs to machine learning models and revealing their responses early on when multiple ``wrong'' inputs are received.
One work that implements such a solution for public cloud-based machine learning APIs is~\citep{chen2019stateful}. 
In it, the authors suggest developing a mechanism to detect and block queries meant to estimate the gradient of the model for an attack that only has query access. 
We believe more research of this nature is needed.
It needs to explore both the specific needs and capabilities of other systems (through careful context-specific threat modeling) and how the adversary's work factor scales when such ``lockout'' defenses are introduced.

\subsection{Robust Input Modalities}
Another fruitful approach we observed for increasing the security of machine learning models is to generate their inputs in a robust manner that increases the work factor for attacks.
While this does not prevent all possible adversarial examples, attackers need to expend many more resources to find adversarial inputs for robust input modalities.

First consider cloud machine learning services where adversaries have the ability to directly alter the pixels that are fed to a computer vision model.
This capability allows the attacker to outright replace the image with one that does not correspond to the world that is being observed and trivially enables serious compromises.
However, in a lot of the systems we observed, adversaries do not have that kind of access and they need to either find ways to project fake objects (as in~\citep{nassi2019mobilbye}) or be more subtle and create physical adversarial objects (as in~\citep{sharif2016accessorize, eykholt2018robust, brown2017adversarial}).
The relative increase in work factor for attackers is large, even though attacks are still possible. 
Projections need specific configurations in order to succeed or expensive equipment.
Adversarial patches require sophisticated knowledge of both machine learning and the model's internals and parameters.
Both of these are much more costly than replacing the raw image input.
This line of thought extends further.
Work on physical adversarial examples in 3D-based vision systems, such as~\citep{cao2019adversarial}, demonstrates the existence of objects that mislead depth-based vision.
However, the objects are both more conspicuous and harder to manufacture. 
Creating them may not be worth the effort for the adversary's payoffs. 
Therefore, we suggest that threat models for AI systems should take into account what the attacker stands to gain from an exploit relative to the resources they need to spend to achieve that goal.

In the course of our experiments with vision systems in particular, we observed that physical adversarial examples exhibit temporal instability.
When the input to the system is video, the predictions of the neural network model exhibit high variance across frames.
In some, the predictions are very close to the adversary's goal.
In others, the adversarial property is lost and the adversary's goal is far from being achieved.
Therefore, to robustify any systems with video input modality in particular, we suggest using temporal inconsistency as a possible signal of an attack in progress.
More generally, combining multiple orthogonal input modalities is likely to increase the effort required to synthesize an adversarial input, and increase the computational burden of the attack as well.
A key research direction is to better understand these costs and the trade-offs between the added burden to the ML developer and system versus the added burden for the attacker.

\subsection{Adaptive Defenses for Adaptive Adversaries}
Some of the most important advances in ML have been driven by common tasks, datasets, and benchmarks.
As long as a model exists that can achieve a good enough benchmark on a given data set, we consider the problem solved.
This has been very useful in core machine learning research and allows us to track and compare progress over time.
It has been so useful that the problem of securing machine learning is often cast as a problem of achieving some level of ``robust accuracy.'' 
To that end, researchers run competitions~\citep{kurakin2018adversarial} where attackers are rewarded for reducing robust accuracy and defenders get points for keeping it high.

There are two problems in applying this approach to developing secure ML systems. 
First, achieving success through a high performance on a benchmark corresponds to a static threat model, in which the capabilities of the attacker are always fixed.
This problem is clearly evidenced by the fact that defenses with high robust accuracy scores have been consistently defeated by adversaries with knowledge of the defense mechanism, as described in~\citep{carlini2017towards, carlini2017adversarial, athalye2018obfuscated, tramer2020adaptive}.
Second, robust accuracy in the face of adversarial examples does not capture the full set of adversarial capabilities available to compromise a model.
Attackers can find out-of-distribution inputs that do not fit in with the inputs that the model designer imagined.
For example, \cite{nassi2019mobilbye} demonstrates an attack that fools an autonomous vehicle into stopping or veering out of its lane by simply projecting false images of stop signs, pedestrians, and lane markers onto and around a road. 
Sufficiently realistic projections of light onto the road are enough to cause the vehicle to violate its security guarantees.
This can often be fixed when a model is retrained with such novel examples or if a dedicated model is developed to reject them.
But adversaries adapt, leading to a whack-a-mole of trying to capture the distribution of ever-changing adversarial inputs.

Indeed, we are not the first to recognize this issue. 
\cite{unrestricted_advex_2018} are running a competition with adaptive adversaries and defenses that focuses on ``unrestricted adversarial examples''---any image where some unambiguous ground truth diverges from the model's high-confidence predictions.
This is a good first step in how the community responds to these threats and we encourage more research in this space so that blind spots in popular models can be thoroughly explored. 
An important research challenge is the automation of such exploration.
Threat modeling $\ell_p$-norm adversarial examples has been made much easier by a well-studied set of algorithms for generating them. 
We challenge the community to explore how new methods, such as simulations or game theory, 
might help in programmatic exploration and reasoning about adaptive adversarial inputs.

\section{Retooling Systems Security for Adversarial Machine Learning}
\label{sec:retooling}
Securing machine learning is a challenging endeavor for two reasons.
First, ML is fundamentally solving a statistical task.
This introduces inherent uncertainty into the input/output behavior of any non-trivial algorithm.
This makes it hard for a defender to have high confidence in evaluating the behavior of ML solutions; is the algorithm simply encountering novel, but real, phenomena, or is it being attacked?
Second, current ML models are very complex and their behaviors are often hard to interpret. 
This lack of insight into their inner workings makes it hard to control these models, resulting in potentially adversarial phenomena like adversarial perturbations.

These challenges are in many ways not unique to ML. 
Modern systems security research deals with subtle, non-deterministic bugs and large, hard-to-interpret code bases as well.
Because the challenges are similar, we believe the goal of secure ML can be helped by an expansion of the focus of systems security research.
In this section, we describe 4 security best practices, what worked well in applying them for ML systems, and where more research is needed to ensure they function well for ML models.
We do not claim to be exhaustive, but the areas we describe below improved the security of ML systems in our explorations.

\subsection{Sanitization of Inputs}
Input sanitization is an important security best practice that identifies invalid inputs that may cause security issues and ensures they are either not processed or stripped of elements that may trigger a flaw.
For instance, in programming languages that allow direct modification of memory, programmers add checks for inputs that may overflow available buffers and reject them if necessary. 
In web applications, form input is stripped of certain characters before being fed into the database engine in order to prevent SQL injection attacks.
Systems with machine learning components should apply similar techniques. 
For example, an autonomous vehicle vision module should not consider painted, printed, or projected images of road signs as real objects to be classified.
For a secure voice command system, a synthesized or recorded voice should be rejected before commands are acted upon.
Similarly, adversarial inputs that are deliberately created to mislead neural networks (e.g., adversarial stickers or patches like the toaster patch from \cite{brown2017adversarial}) should be considered invalid inputs.
Thus, in a real system, processing of these kinds of inputs should be stopped or ignored in order to ensure secure execution.

Unfortunately, solutions to this challenge have received less attention in the research literature.
As a result, we found that methods for ensuring only valid inputs reach machine learning models are less robust and more ad-hoc than the models themselves.
A key challenge is that input validity itself is ill-defined for many systems and applications.  
For many use cases, no simple heuristics exist to reject invalid inputs, or to identify safe behavior or response in the face of an invalid input.

Even when input validity is well-defined, ML models for sanitizing inputs are brittle and unreliable.
This is partially a problem of datasets.
Consider computer vision.
Large, labeled datasets exist for a diverse set of tasks, such as image classification and object detection. 
However, none of these labels printed or projected objects distinctly from ``naturally occurring'' ones. 
Thus, it is to be expected that object detectors detect both printed and projected objects as well as real ones.
Similarly, in audio transcription, the benchmark dataset for detecting spoofed recordings contains only 18,000 utterances~\citep{kinnunen2017asvspoof}, as compared to over 65,000 utterances in the transcription task dataset~\citep{warden2018speech}.
This indicates that datasets with adversarial inputs may be too small to solve the task well with any model.
Adversaries can draw on a much larger space of possibly adversarial inputs than those that are captured in the data.
To remedy this, we call for the creation of larger and more diverse input sanitization datasets that are continuously updated with new adversarial behavior.

However, even if good datasets existed, the models trained on them are also likely to be vulnerable to adversarial examples.
Thus, we believe more research is needed in this area.
A primary question to answer is whether adversarial examples and adversarial physical objects can be created that fool sanitization models as well. 
More importantly, research should seek to understand if one can generate inputs that are adversarial to both kinds of models at the same time. 
A related question is whether more models working together increases the adversary's work factor. 
Finally, an alternative approach to input sanitization is anomaly detection and outlier rejection.
It is currently unknown which approach is likely to yield better, more generalizable results and we believe research should work to address this gap of knowledge.

\subsection{Evaluating and Testing for Security}
A well-agreed security principle is that the designers of a system should not be solely responsible for its security evaluation.
Independent review serves to catch vulnerabilities that designers would have missed otherwise. 
There are two schools of thought on who should aid them in doing so.
Regardless of the position one takes on this, systems security for ML needs better solutions for both.

On the one hand, many security experts believe that details of the system internals should be made available publicly to allow for better scrutiny.
This not only allows for more people to look at the design and implementation and look for security flaws but also forces designers to create software that is secure even if those details are known to the attacker. 
Since even the most well-guarded secrets leak, it is believed that this makes more secure software overall.
Unfortunately, releasing a model's architecture and parameters enables strictly stronger attacks in the ML world.
This is due to the fact that backpropagating through a neural network (when adversaries possess its weights and structure) remains the most efficient way to generate adversarial examples for it.
Query-access methods generally take many more queries, as the gradient function needs to be approximated and cannot be analytically computed.
This is also born out by our observation that designers of ML systems prefer to rely on obfuscating the design of their system.
We believe more research is needed on how end-to-end ML systems should be designed so that they are secure even if the ML model and other details of it become public.

On the other hand, not all secure systems are public.
Private auditing and internal testing serve to discover and remedy flaws in those cases.
This is generally aided by good software engineering practices that isolate functionality and make individual components easy to interpret.
However, even the smallest unit of functionality in ML systems --- the model itself --- is often very complex.
Thus, it is difficult to reason about corner cases of isolated components and test for them explicitly.
Early research has already appeared on how this can be achieved programmatically.
For example,~\cite{pei2017deepxplore}, developed a framework for automated testing guided by ``neuron coverage,'' a concept stemming from software code coverage.
Indeed, many works have followed in this paper's footsteps.
Without claiming to be exhaustive, we highlight works such as~\citep{ma2018deepgauge, wang2018efficient, odena2018tensorfuzz} and encourage further research in this area.

\vspace{-10pt}
\subsection{Updates}
No system is perfectly secure and vulnerabilities are often discovered well after software is first released to users.
The classical recommendation to deal with this issue in systems security is to allow for continuous updating.
This allows engineers to release patches as soon as a fix is available and prevents exploitable vulnerabilities from languishing in deployed software.
Certainly, ML models often need updating as well.
They operate on phenomena that may exhibit distributional shift. 
New situations may arise that the model has not seen, leading to degraded performance and raising the risk of exploits due to unanticipated behavior.
In other cases, developers might learn that certain classes of inputs cause the model to fail consistently (such as projected images on the road for object detection). 
In those situations, it is highly desirable for engineers to release new versions of the model trained on updated data or that are otherwise more secure.

However, ML models often introduce system dependencies that make such updating difficult.
We observed particularly troublesome cases in pipelined settings where  the code processing the outputs of the model expects them to be distributed in specific ways.
As a concrete example, systems that process stored data generated by machine learning models need to keep old versions of models running in parallel to newer ones until enough data processed by the new version has been generated. 
Adapting the code is also not an option until enough new data exists.
This slows down the update cycle and prevents engineers from issuing ``hot'' patches to remedy newly discovered security issues. 
Therefore, exploitable security vulnerabilities in machine learning systems due to the model often persist for long periods of time.

One concrete suggestion is to decouple the classification task that the model performs from the task of securing input to it.
Techniques such as adversarial training attempt to directly robustify the classification model itself.
We found that keeping the model itself unchanged, but guarding it with a set of defense algorithms is a more scalable and adaptive solution.
This allows for faster deployment of ``hot'' patches when new attacks are found.

A more thorough discussion of these issues from the perspective of software engineering is available in~\citep{sculley2014machine} but we highlight the problems with updates that we observed in the security setting.
We urge the research community to consider the issue of updating models so that they can output data that conforms to a standard specification.
Any solutions should allow developers to release new models that are backwards compatible with previously stored outputs.
However, as in all cases of backward compatibility, such research needs to balance future performance and security with conforming to the old standard.

\subsection{Sharing Knowledge of Vulnerabilities}

Because machine learning models are needed to let systems process real-world phenomena, it is nearly impossible to predict a-priori all the ways in which the distributions for certain tasks can shift. 
Motivated adversaries are likely to keep discovering ways of violating the assumptions of the training set and the system more broadly. 
Thus, system designers should be ready to handle failure cases as they arise.
That is why we propose that the community establish a common database of failure modes.
If practitioners share with each other where models have failed and provide examples, then they can collaborate in covering a large search space of potential security holes. 
Indeed, this concept is not new; the CVE database of public security vulnerabilities in classic software (\url{https://cve.mitre.org}) has long served as a centralized repository of known exploits that developers can reference to secure their own code and anticipate threats. 

We believe that the machine learning community can benefit from a similar database.
For example, all researchers and practitioners studying object detection can submit videos where important objects are missed or where spurious objects are detected.
If this database is made public and labels are provided in standardized formats, everybody applying detectors to their problem can use the data directly to fit a better function that reduces false positives and false negatives.
But even the experience of observing new \textit{classes} of failure cases can prod systems designers to remedy problems in their vision pipeline.
Each practitioner can go out and collect extra data to augment their model or they could build in heuristics to better handle inputs with the newly observed failures. 
In either case, such a public database would help defenders by reducing the number of ``unknown unknwons'' that designers have to deal with in favor of ``known unknowns.''

\section{Discussion and Conclusions}

Machine learning introduces novel challenges to securing computer systems.
Not only are models applied in statistical settings where it is hard to anticipate adversarial input, but even the most sophisticated models contain vulnerabilities that we do not know how to remedy yet.
We argue that these deficiencies can be filled by learning from experiences with deployed systems using ML and carefully reasoning through how closely they follow classical security best practices. 
This requires two advances.
On the one hand, machine learning needs to adapt to consider full systems and the real threats they are likely to face in an adaptive setting.
On the other, technological solutions for applying systems security best practices in ML systems need to be developed.

In order to remedy acute threats against deployed systems, machine learning practitioners should consider these approaches:
\begin{itemize}[leftmargin=*, itemsep=1pt]
    \item Threat model for adaptive adversaries who possess knowledge of your defenses.
    \item Consider how exploits in ML translate to exploits of the overall system and examine whether the programming logic advantages attackers in adversarial settings.
    \item Increase the cost of an attack by introducing multiple models that sanity-check each other.
    \item Introduce robust input modalities where adversarial inputs are inherently more costly.
\end{itemize}

However, these mitigations are only likely to last so long.
In order to move the field of secure ML forward, we call on researchers to provide solutions to the following problems plaguing real systems:
\begin{itemize}[leftmargin=*,itemsep=1pt]
    \item Provide mechanisms for rejecting invalid and outright adversarial inputs so that machine learning models do not process those and enable exploits.
    \item Develop methods to easily examine machine learning models for security pitfalls in testing.
    \item Propose models that can safely be released publicly and do not increase the adversary's capabilities at a disproportionate cost to the benefit from open access
    \item Study update mechanisms that preserve ``good'' functionality and the distribution of outputs so that vulnerable machine learning models can be quickly updated.
    \item Compile a standardized database of common failure cases that ML practitioners can benefit from in designing their systems around known vulnerabilities.
\end{itemize}

We are optimistic that these changes can allow us to make better and more reliable use of the impressive functionality enabled by machine learning today.

\section*{Broader Impact}

The work presented here is intended to benefit designers of ML systems, researchers who study issues around security and AI, and -- ultimately -- end users who should benefit from more secure and reliable software. 
Unfortunately, ML is not strictly used for ``good'' and where these principles might aid in securing systems with benign functionality, they might also make it hard to develop defenses against malign ML systems (such as face recognition for illegitimate surveillance) if they are robustified in these ways. 
However, we believe that increasing the security of software is a goal in and of itself and will ultimately have many more benefits than harms.
We do not believe that issues of bias in datasets apply, as we do not use any datasets in this work. 

\begin{ack}
We thank Ofir Press, Ludwig Schmidt, Pascal Sturmfels, and Jacob Steinhardt, for helpful discussions and their thoughtful feedback on this paper. 
This research was supported in part by the University of Washington Tech Policy Lab, which receives support from: the William and Flora Hewlett Foundation, the John D. and Catherine T. MacArthur Foundation, Microsoft, the Pierre and Pamela Omidyar Fund at the Silicon Valley Community Foundation.
\end{ack}



\bibliographystyle{plainnat}
\bibliography{example_paper}

\end{document}